\documentclass[prl,reprint,twocolumn,superscriptaddress,showpacs]{revtex4-1}

\usepackage{graphicx}
\usepackage{epsfig}
\usepackage{amsmath,bbm,amssymb,bm,times}
\usepackage[colorlinks,linkcolor=blue,citecolor=blue]{hyperref}

\newcommand{\eqn}[1]{
\begin{eqnarray}
	#1
\end{eqnarray}
}

\begin{document}
\title{Quantum Rydberg Central Spin Model}
\author{Yuto Ashida}
\email{ashida@ap.t.u-tokyo.ac.jp}
\affiliation{Department of Applied Physics, University of Tokyo, 7-3-1 Hongo, Bunkyo-ku, Tokyo 113-8656, Japan}
\affiliation{Department of Physics, University of Tokyo,  7-3-1 Hongo, Bunkyo-ku, Tokyo 113-0033, Japan}
\author{Tao Shi} 
\email{tshi@itp.ac.cn}
\affiliation{CAS Key Laboratory of Theoretical Physics, Chinese Academy of Sciences, Beijing 100190, China}
\author{Richard Schmidt}
\affiliation{Max-Planck-Institut f\"ur Quantenoptik, Hans-Kopfermann-Strasse. 1, 85748 Garching, Germany}
\affiliation{Munich Center for Quantum Science and Technology (MCQST), Schellingstr. 4, 80799 M\"unchen, Germany}
\author{H. R. Sadeghpour}
\affiliation{ITAMP, Harvard-Smithsonian Center for Astrophysics, Cambridge, MA 02138, USA}
\author{J. Ignacio Cirac}
\affiliation{Max-Planck-Institut f\"ur Quantenoptik, Hans-Kopfermann-Strasse. 1, 85748 Garching, Germany}
\affiliation{Munich Center for Quantum Science and Technology (MCQST), Schellingstr. 4, 80799 M\"unchen, Germany}
\author{Eugene Demler}
\affiliation{Department of Physics, Harvard University, Cambridge, MA 02138, USA}

\date{\today}
\begin{abstract} 
We consider dynamics of a Rydberg impurity in a cloud of ultracold bosonic atoms in which the Rydberg electron can undergo spin-changing collisions with surrounding atoms. This system realizes a new type of the quantum impurity problem that compounds essential features of the Kondo model, the Bose polaron, and the central spin model. To capture the interplay of the Rydberg-electron spin dynamics and the orbital motion of atoms, we employ a new variational method that combines an impurity-decoupling transformation with a Gaussian ansatz for the bath particles. We find several unexpected features of this model that are not present in traditional impurity problems, including interaction-induced renormalization of the absorption spectrum that eludes simple explanations from molecular bound states, and long-lasting oscillations of the Rydberg-electron spin. We discuss generalizations of our analysis to other systems in atomic physics and quantum chemistry, where an electron excitation of high orbital quantum number interacts with a spinful quantum bath.  
\end{abstract}

\maketitle
Many important phenomena in strongly correlated many-body systems can be understood from the perspective of three fundamental systems: the Kondo impurity model, the Bose polaron model, and the central spin problem.  
{\it The Kondo model } \cite{KJ64} is characterized by the breakdown of perturbation theory due to the effective enhancement of the antiferromagnetic interaction and has played a vital role in understandings of heavy fermion materials \cite{LH07,*SQ10} and mesoscopic structures \cite{JTD_book}. Considerable theoretical effort has also been invested in understanding the Bose Kondo problem \cite{FGM04,*FS06,*FFM11,*FT15}, which was argued to emerge at the transition point between the antiferromagnetic  and the paramagnetic phases \cite{ZL04}. 
 {\it The Bose polaron model} has been introduced by Landau, Pekar \cite{Pekar} and Fr{\"o}hlich \cite{Frohlich52}. The concept of polaronic dressing was applied to a broad range of systems, where a mobile particle interacts with a bath of collective modes. In particular, dynamics of charge carriers in doped antiferromagnetic Mott insulators, such as high $T_c$ cuprates, can be understood from the perspective of magnetic polarons  \cite{JTD85,*OR00,*FB11,*KoM12,*RS12,*EDK17,*SM17}.  
 Recently, Bose polarons have been actively explored also in ultracold atoms \cite{TJ09,NA10,CW11,*CW112,RS13,VJ13,LW14,SYE162,Levinsen2015,Giorgini2015,*Giorgini2016,CRS15,YA17,HMG16,JNB16,PLA18,YZZ19}.

\begin{figure}[b]
\includegraphics[width=70mm]{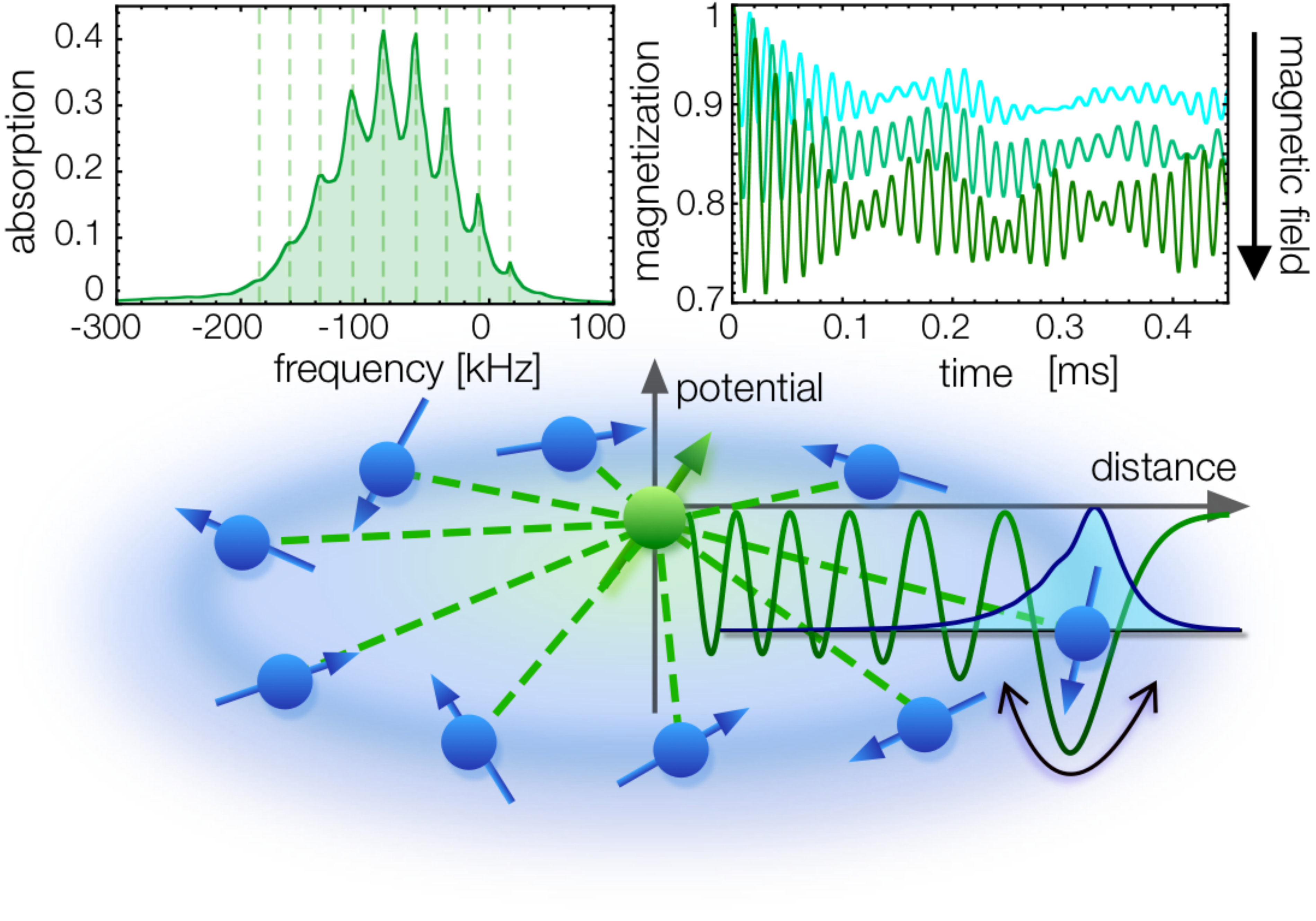} 
\caption{\label{fig1}
A photoexcited Rydberg electron undergoes spin-changing collisions in a cloud of ultracold atoms. The interaction between the Rydberg electron and bosonic atoms includes both motional and Kondo-type components.  Many-body effect manifests itself as interaction-induced renormalization of the absorption spectrum  (top left). The central spin exhibits long-lasting oscillations that strongly depend on the external magnetic field (top right). 
}
\end{figure}

 While these two canonical models deal with {\it delocalized} environmental modes that can move in space, {\it the central spin model} describes a single (central) two-level system nonlocally coupled to {\it localized} modes such as nuclear spins  \cite{MG76,NVP00,TJM03,DJ04,EAY05,BM07,WZ07,CG07,LB08,BM10,WWM10,WWM12,KEM12,TAM12,FA13,RDA18}. Experimental systems described by the central spin model include quantum dots \cite{KA03,*JS03,*CWA04}, superconducting-flux qubits \cite{CI03}, and nitrogen-vacancy centers in diamond \cite{HR08}.  
 The notable feature of this model is its integrability, i.e., dynamics is strongly constrained by the extensive number of integrals of motion, leading to long-lived coherent oscillations and  formation of solitons.
Altogether, the above three paradigmatic classes of many-body systems exhibit distinct physics and have so far been studied individually in different contexts.

The aim of this Letter is to propose and analyze a new type of quantum many-body problems unifying the above distinct paradigms. 
The key element is a Rydberg excitation in a cloud of ultracold atoms, which  undergoes spin-changing collisions with surrounding atoms (see Fig.~\ref{fig1}). The spin of the Rydberg electron plays the role of the central spin that interacts with mobile environmental bosons via ultralong-range Kondo couplings.  When the  Rydberg-electron spin is flipped, a single bath spin is also flipped due to the spin conservation while the scattering potential for the rest of atoms is strongly altered. In turn, there is a feedback from orbital motion of atoms on spin dynamics since the spin interaction depends on atomic positions.  From now on, we will refer to this class of systems as the Rydberg Central Spin Model (RCSM). To solve this  challenging problem, we develop a new theoretical approach that combines a recently proposed impurity-decoupling transformation with the variational Gaussian ansatz for bosons. 
We make several concrete predictions that can be tested by current experimental techniques. 

One of the key features of the RCSM is formation of many-body bound states, which manifests itself as interaction-induced renormalization of peaks in the absorption spectrum, which eludes simple explanations based on molecular states.   
This should be contrasted to earlier studies of Rydberg spectroscopy that were either performed in the low-density regime, where many-body aspects were not important \cite{ADA14,SH15,BF16,MD19,EF19}, or did not involve spin-changing collisions  \cite{GC00,ELH02,BV09,GA14,DBJ15,RS16,SM16,CF18,*RS18,KKS18}.  Another surprising feature is long-lasting oscillations of the central spin. Such oscillations are absent in the infinite-mass limit of bath particles, where the system reduces to the conventional central spin problem, revealing the crucial role of the orbital dynamics of environmental atoms. Furthermore, we find that the oscillation frequency has a nonanalytic dependence on the density of environmental atoms, characteristic of nonperturbative many-body dynamics.  These results demonstrate that the RCSM is fundamentally distinct from both the usual central spin model \cite{MG76,NVP00,TJM03,DJ04,EAY05,BM07,WZ07,CG07,LB08,BM10,WWM12,KEM12,FA13,RDA18} and the previously studied (spinless) Rydberg Bose polaron  \cite{RS16,SM16,CF18,*RS18,KKS18}.

{\it Rydberg Central Spin Model.---} We consider a Rydberg impurity interacting with a spinful bosonic  environment of particle density $\rho$. The Rydberg impurity has an electron with a high principal quantum number $n$ and orbital wavefunction $\Psi_{e}({\bf r})$, whose size can surpass the average interparticle distance $\rho^{-1/3}$.
We consider the situation when only two internal states of environmental bosons need to be included. Our first goal is to introduce and analyze the simplest setup of the RCSM. Thus, for now we assume that the interaction between the Rydberg electron and bosons exhibits SU(2) symmetry. While this symmetry may be lost when the full algebra of angular momentum is included as discussed later, this will not change any results substantially, i.e., most phenomena  described below are generic features of the RCSM. 

The interaction between environmental bosons and the Rydberg electron is given by  Fermi's pseudopotential \cite{EF34}: $V_{{\rm T},{\rm S}}({\bf r})\!=\!2\pi\hbar^{2}a_{{\rm T},{\rm S}}|\Psi_{e}({\bf r})|^{2}/m_{e}$, where $a_{{\rm T},{\rm S}}$ are the zero-energy triplet (T) and singlet (S) scattering lengths and $m_e$ is the electron mass. 
The long-range interaction between the Rydberg impurity and the surrounding  bosons is then described by $V_{{\rm T}}\hat{P}_{{\rm T}}\!+\!V_{{\rm S}}\hat{P}_{{\rm S}}$, where $\hat{P}_{{\rm T}}\!=\!\hat{{\bf S}}_{e}\cdot\hat{{\bf S}}_{{\bf r}}\!+\!3/4$ and $\hat{P}_{{\rm S}}\!=\!1\!-\!\hat{P}_{{\rm T}}$ are the projection operators onto the triplet and singlet channels. Here, $\hat{{\bf S}}_{e}\!=\!\hat{\boldsymbol{\sigma}}_{e}/2$ and $\hat{{\bf S}}_{{\bf r}}\!=\!\sum_{\alpha\beta}\hat{\Psi}_{{\bf r}\alpha}^{\dagger}(\boldsymbol{\sigma}/2)_{\alpha\beta}\hat{\Psi}_{{\bf r}\beta}
$ are the spin operators of the Rydberg electron and environmental atoms, respectively, with $\hat{\Psi}^\dagger_{{\bf r}\alpha}$ ($\hat{\Psi}_{{\bf r}\alpha}$) being the bosonic creation (annihilation) operator at position $\bf r$ with internal state $\alpha\!=\Uparrow,\Downarrow$. 

The  total system is thus governed by the Hamiltonian
\eqn{\label{Hamiltonian}
\hat{H} =  \hat{H}_{0}+\hat{{\bf S}}_{e}\cdot\int d{\bf r}g_{{\bf r}}\hat{{\bf S}}_{{\bf r}}+h_{z}\hat{S}_{e}^{z},
}
where $\hat{H}_{0}\!=\!\sum_{\alpha}\int d{\bf r}\hat{\Psi}_{{\bf r}\alpha}^{\dagger}{h}_{0}\hat{\Psi}_{{\bf r}\alpha}$ is the quadratic part with ${h}_{0}\!=\!-\hbar^{2}\nabla^{2}/(2m)\!+\!(3V_{{\rm T}}+V_{{\rm S}})/4$ and $m$ being the mass of environmental bosons. The second term describes the interaction between the Rydberg central spin and the bath with long-range Kondo couplings $g_{{\bf r}}\!=\!V_{{\rm T}}\!-\!V_{{\rm S}}$. The magnetic field $h_z$ should be understood as the difference in the Zeeman energies of the impurity and bath spins  due to different $g$-factors 
\footnote{We note that the total $S^z_{\rm tot} $ is conserved in Eq.~(\ref{Hamiltonian}) and thus, the term proportional to it has been omitted.}. We neglect the boson-boson interaction since the Rydberg potentials have considerably larger energy scales. We focus on the Rydberg electron with zero angular momentum.

We consider a sudden quench of the Rydberg interactions starting from the initial state 
\eqn{\label{initial}
|\Psi_{0}\rangle=|\!\uparrow\rangle_{e}|{\rm BEC}_{\Downarrow}\rangle,
}
where $|\!\!\uparrow\rangle_e$ is the spin-up state of the Rydberg electron and $|{\rm BEC}_{\Downarrow}\rangle$ is the zero-temperature Bose-Einstein Condensate (BEC) of environmental atoms polarized in the $\Downarrow$ state. This quench corresponds to photoexciting an electron from the ground state to the excited Rydberg  state. 
The spectral function measured experimentally is given as \cite{RS16}: $A(\omega)\!=\!{\rm Re}[\int_{0}^{\infty}dt e^{i\omega t}S(t)]$ with $S(t)\!=\!\langle \Psi_0|e^{-i\hat{H}t/\hbar}|\Psi_0\rangle$ \footnote{We here do not include the contribution from the free time evolution without Rydberg interactions since it merely shifts the absorption spectrum by a global trivial constant.}. We will also analyze the impurity magnetization $m_z(t)\!=\!\langle\hat{\sigma}^z_e(t)\rangle$.

{\it Variational approach with the impurity decoupling.---} 
The many-body problem \eqref{Hamiltonian} presents a new class of condensed matter models, creating a formidable theoretical challenge; one has to solve the full many-body evolution by taking into account  the impurity-environment entanglement mediated by the central spin couplings as well as the orbital motion of environmental particles.  We tackle this challenge with a new variational approach based on an impurity-decoupling transformation. 
The key idea is to utilize parity symmetry of the Hamiltonian (\ref{Hamiltonian}) to decouple the impurity spin degree of freedom. The parity symmetry corresponds to the $\pi$ rotation around $z$ axis and is given by $\hat{{\rm P}}=\hat{\sigma}_{e}^{z}e^{i\pi\hat{N}_{\Uparrow}}$, where $N_\Uparrow$ is the number of spin-up environmental bosons. The operator $\hat{{\rm P}}$ has eigenvalues $\pm 1$, so it does not come as a surprise that there is a unitary transformation 
$\hat{U}\!=\!(1\!+\!i\hat{\sigma}_{e}^{y}e^{i\pi\hat{N}_{\Uparrow}})/\sqrt{2}$, which maps it into the impurity spin \cite{YA18L,*YA18B}: 
\eqn{
\hat{U}^{\dagger}\hat{{\rm P}}\hat{U}=\hat{\sigma}_{e}^{x}.
} 
Since the initial state~\eqref{initial} resides in the sector $\hat{\rm P}\!=\!+1$,  the  time evolution can be described by the transformed Hamiltonian 
$\tilde{\hat{H}}= \hat{U}^{\dagger}  \hat{H} \hat{U} $
conditioned on a classical variable $\hat{\sigma}_{e}^{x}\!=\!+1$, where only the environmental degrees of freedom contribute to dynamics.  In this decoupled frame, we approximate the environmental state by a bosonic Gaussian state \cite{WC12} and employ the time-dependent variational principle \cite{JR79,KP08,ST17} to analyze the out-of-equilibrium dynamics.

\begin{figure}[t]
\includegraphics[width=80mm]{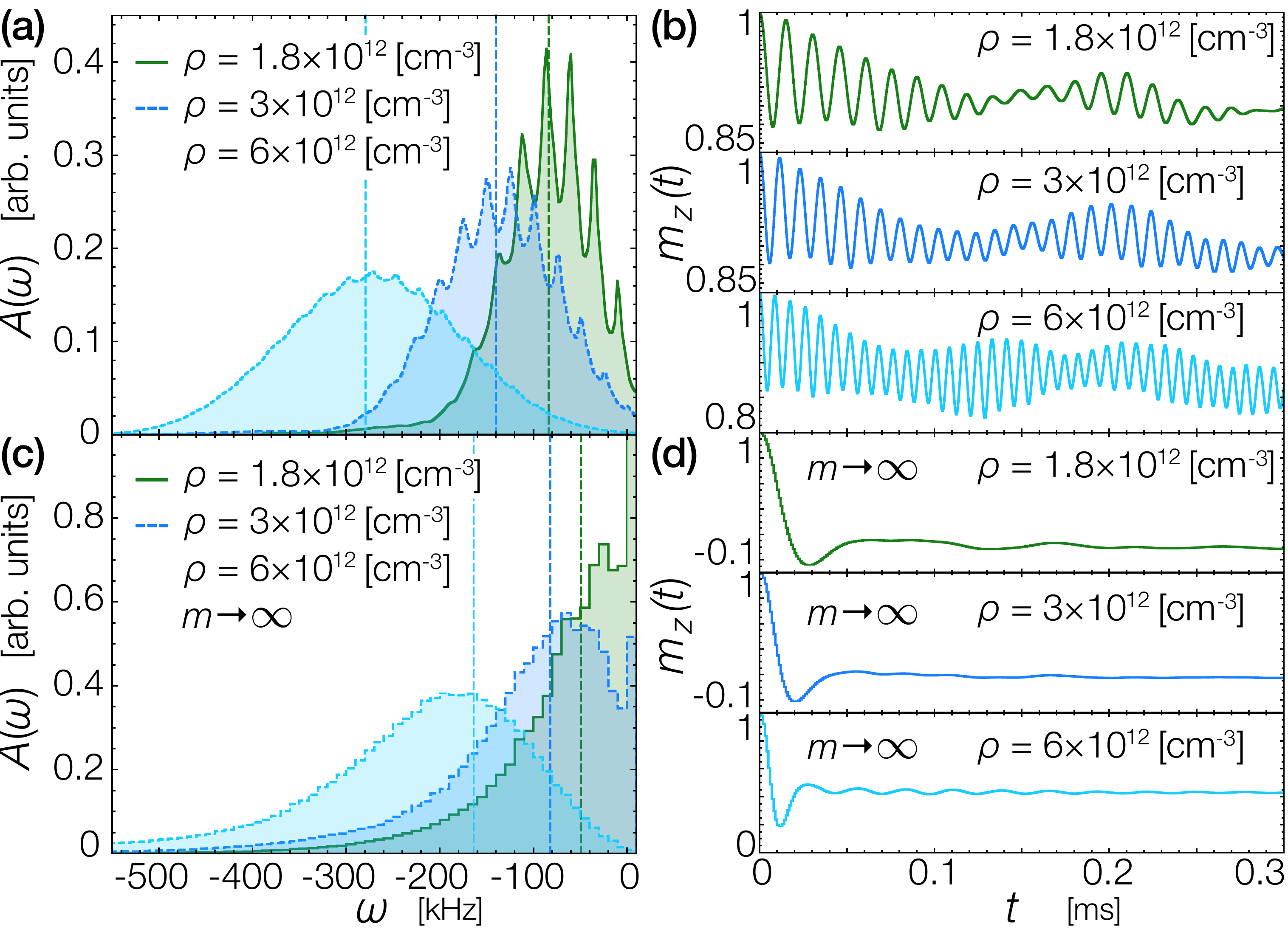} 
\caption{\label{fig2}
(a,c) Absorption spectra $A(\omega)$ at different densities $\rho$ of (a) mobile and (c) immobile environmental atoms. Dashed lines indicate the mean-field shifts $\Delta_{\rm MF}$ of the spectra. (b,d) Central spin dynamics $m_z(t)\!=\!\langle\hat{\sigma}_e(t)\rangle$ after a quench with (b) mobile and (d) immobile environmental spins. We set the magnetic field to be zero $h_z=0$. 
}
\end{figure}

{\it Results.---} 
Our main goal is to reveal generic nonequilibrium features of  the RCSM rather than to make predictions specific to particular experimental setups. To this end, we use a potential profile created by an excited electron of $^{87}{\rm Rb}(87s)$ scattering with $^{87}{\rm Rb}$ atoms as a typical example of Rydberg potentials. However,  qualitative features of the dynamics in the RCSM are insensitive to details of the potentials.

Figure~\ref{fig2}a shows the results for the absorption spectra $A(\omega)$ at different densities $\rho$.  Details of the analysis are presented in the companion paper \cite{YA19A}. With increasing density, the spectra acquire a Gaussian-shape and their centers move to larger detunings. As indicated by the vertical dashed lines, we find that these detunings are consistent with the mean-field (MF) shifts $\Delta_{{\rm MF}}\!=\!\langle\Psi_{0}|\hat{H}_{\parallel}|\Psi_{0}\rangle\!\propto\!\rho$ of the Hamiltonian with the longitudinal coupling $\hat{H}_{\parallel}\!=\!\hat{H}_{0}\!+\!\hat{S}_{e}^{z}\int d{\bf r}g_{{\bf r}}\hat{S}_{{\bf r}}^{z}$.
Using $\hat{S}_e^z\!=\!+1/2$, it reduces to a noninteracting quadratic Hamiltonian with the mean Rydberg potential $V_{{\rm mean}}=V_{0}-g_{{\bf r}}/4=(V_{{\rm T}}+V_{{\rm S}})/2$. These facts indicate that, at the level of this mean-field feature, the flip-flop interaction $\hat{H}_{\perp}\!=\!\int d{\bf r}g_{{\bf r}}(\hat{S}_{e}^{-}\hat{S}_{{\bf r}}^{+}\!+\!{\rm h.c.})/2$ does not play a significant role as consistent with a largely polarized central spin (cf. Fig.~\ref{fig2}b).

\begin{figure}[b]
\includegraphics[width=83mm]{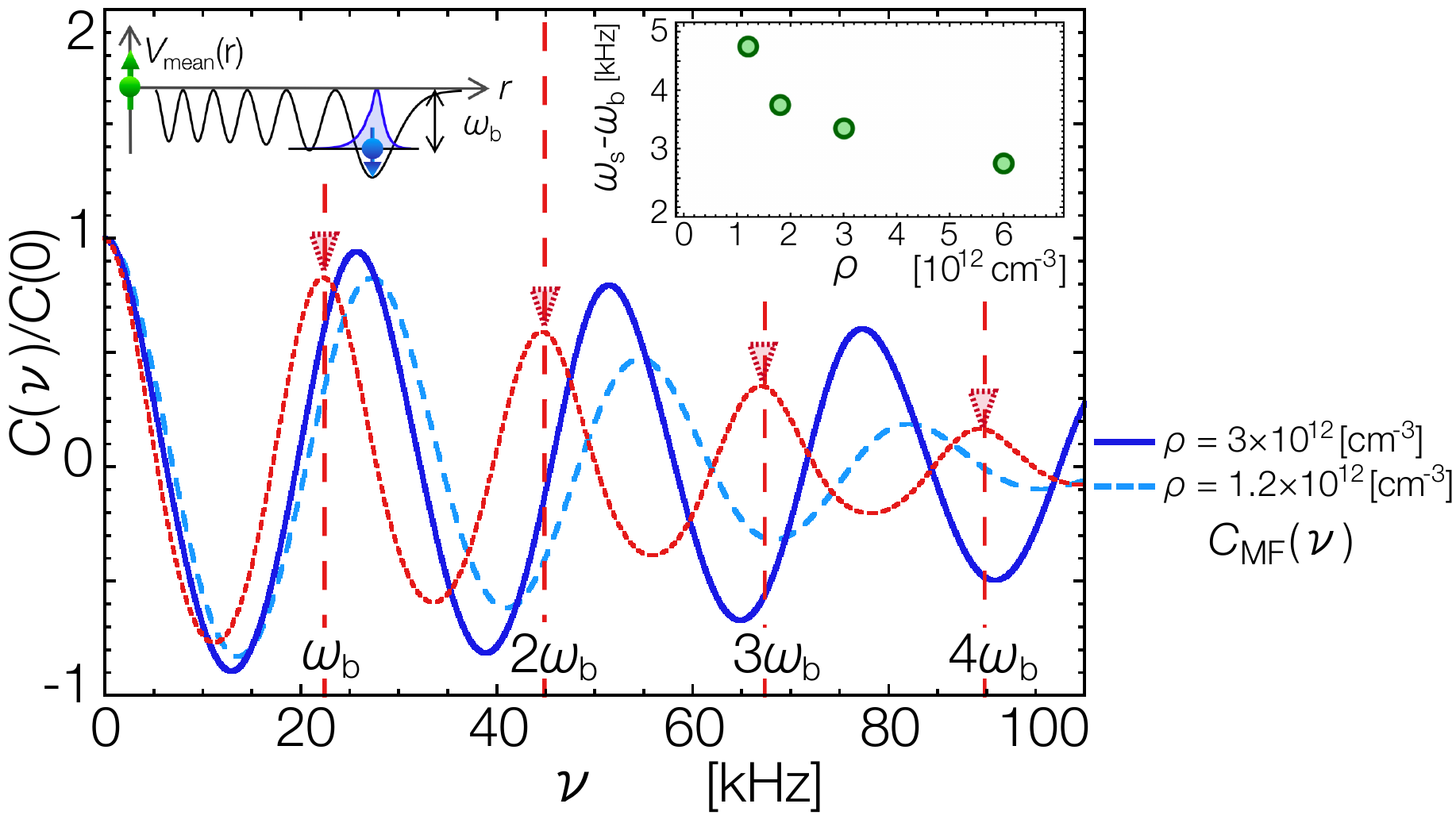} 
\caption{\label{fig3}
Correlation $C(\nu)$ of the absorption spectrum with detuning $\nu$  (main panel). The blue solid and dashed curves (red dotted curve) show the results obtained by quenching the full interacting Hamiltonian $\hat{H}$ (the quadratic Hamiltonian $\hat{H}_\parallel$).  The red dashed vertical lines indicate multiple values of the dominant bound-state energy $\omega_{\rm b}$ of $V_{\rm mean}$ (cf. left inset), which match with the peak positions of the quadratic result. The many-body features appear as deviations of the interacting results from the single-particle energies and their sensitivity to environmental density (cf. right inset). We set $h_z=0$. 
}
\end{figure}

In contrast, while the existence of peaks in Fig.~\ref{fig2}a comes from molecular physics, the interaction-induced renormalization of their positions and spacings has a many-body origin intrinsic to the RCSM. To show this, in Fig.~\ref{fig3} we plot   the correlation function of the spectrum $C(\nu)\!=\!\int d\omega\delta A(\omega)\delta A(\omega\!+\!\nu)$  with detuning $\nu$, where $\delta A(\omega)$ denotes the absorption spectrum subtracted from a fitted Gaussian profile.  
 For comparison, we also present $C_{\rm MF}(\nu)$ obtained using the quadratic Hamiltonian $\hat{H}_\parallel$ (red dotted curve), where $V_{{\rm mean}}$ characterizes the mean-field potential experienced by environmental bosons in a high-density limit. 
 The maximal values of $C_{\rm MF}$ correspond to  integer multiples  of the single-particle energy $\omega_{\rm b}$ of the dominant bound state localized in the outermost well of  $V_{{\rm mean}}$ (cf. left inset in Fig.~\ref{fig3}). Corresponding energies are thus independent of environmental density and determined by the two-body problem. In contrast, the blue solid and dashed curves show the results for $C(\nu)$ corresponding to the quench of the full interacting Hamiltonian $\hat{H}\!=\!\hat{H}_\parallel\!+\!\hat{H}_\perp$ and exhibit much richer structures. 
The many-body nature of the resolved peaks in $A(\omega)$ manifests itself in the departure of the peak-spacing frequency $\omega_{\rm s}$ from $\omega_{\rm b}$ and also in its sensitivity to environmental density (cf. right inset in Fig.~\ref{fig3}). 
Its convergence in the mean-field limit $\rho\to\infty$ is slower than the scaling $\propto 1/\rho$ characteristic of the standard central spin problem \cite{KA03}. In a very low-density regime $\rho<10^{12}{\rm cm}^{-3}$, the resolved peaks converge to bound-state energies for the triplet and singlet Rydberg potentials and lose the equal-spacing feature.
Our findings originate from the Kondo-type dressing of the bare Rydberg molecules and go beyond the previously analyzed cases of  quadratic Hamiltonians \cite{GC00,ELH02,BV09,GA14,DBJ15,RS16,SM16,CF18,*RS18,KKS18}. The results remain qualitatively the same as long as the scattering-length difference is large such that the spin-exchange interaction $\hat{H}_{\perp}$ plays a significant role.  

\begin{figure}[t]
\includegraphics[width=57mm]{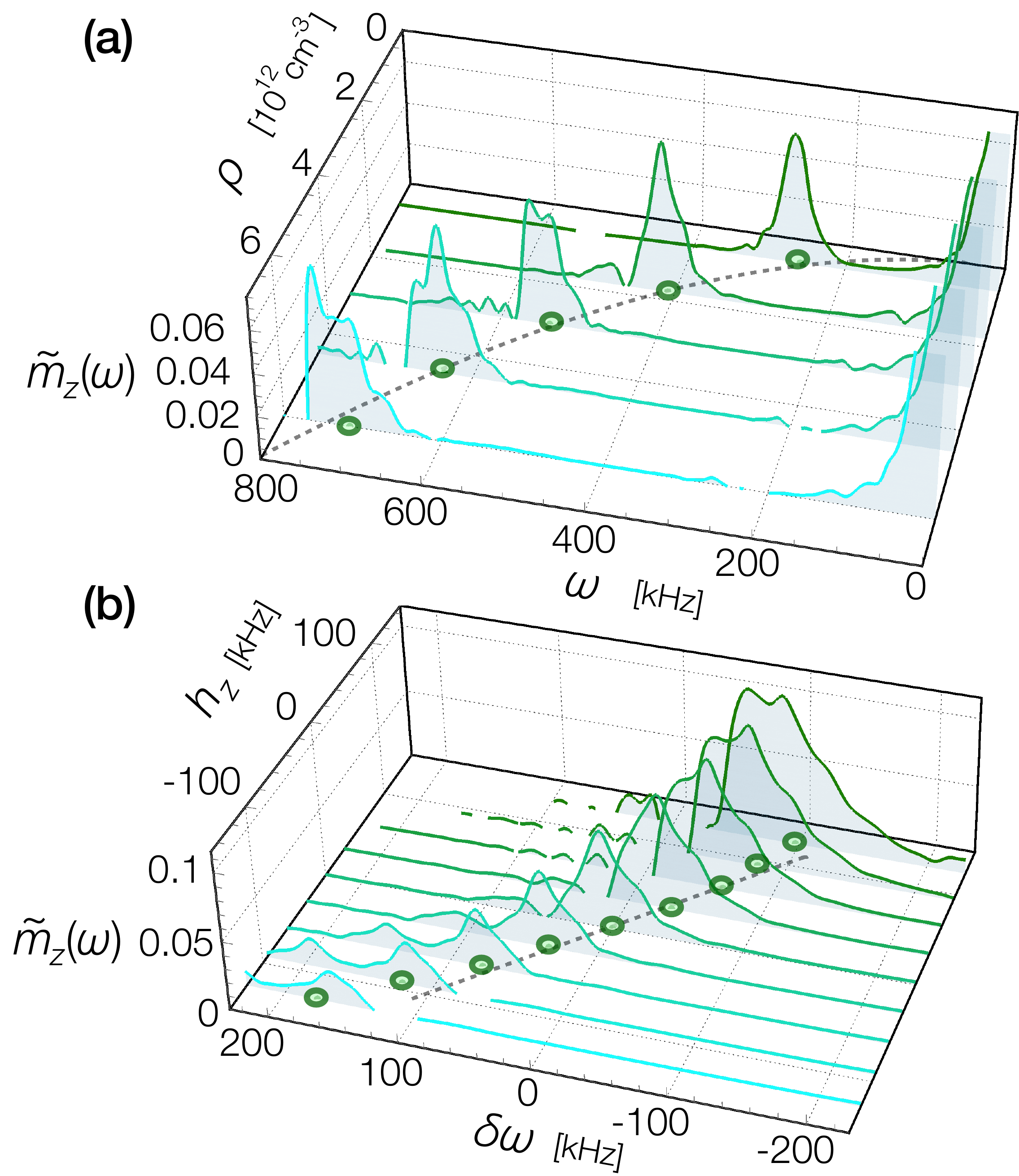} 
\caption{\label{fig4}
The Fourier spectra $\tilde{m}_{z}(\omega)$ of the central spin dynamics $m_z(t)$  (a) at different particle densities $\rho$ with a zero magnetic field and (b) at different magnetic fields $h_z$ with $\rho\!=\!1.8\!\times \!10^{12}$~cm$^{-3}$. The black dashed curve and line at the bottom planes indicate the square root scaling $\omega\!\propto\!\sqrt{\rho}$ in (a) and the linear relation $\delta\omega\!=\! -h_z$ in (b), respectively. The circles at the bottom planes indicate the mean frequencies of the spectra around the peak values. 
}
\end{figure}

Figure~\ref{fig2}b shows the corresponding central spin dynamics $m_z(t)\!=\!\langle\hat{\sigma}^z_e(t)\rangle$. Firstly, the nondecaying magnetization is one of the key features of the central spin problem with an initially fully polarized environment \cite{BM07};  only a small portion of a many-body state with the opposite central spin $|\!\downarrow\rangle_e$ can be admixed due to a large energy cost to flip the central spin immersed in a polarized environment. Secondly, the Rydberg spin exhibits long-lasting oscillations whose frequency $\omega_{\rm mag}$ increases with higher densities. To further investigate the dependence of $\omega_{\rm mag}$ on density $\rho$, we plot in Fig.~\ref{fig4}a  the Fourier spectra $\tilde{m}_{z}(\omega)$ of  the  dynamics $m_z(t)$. As inferred from the black dashed curve at the bottom of the plot, we find a square root scaling  $\omega_{\rm mag}\!\propto\!\sqrt{\rho}$ that is dramatically different from the conventional linear scaling found in studies of the ordinary central spin problem \cite{KA03,*JS03,*CWA04}. The nonanalytic behavior implies that a nonperturbative treatment (as performed here) is essential for the analysis of the RCSM. 

The oscillation frequency and amplitude of the central spin can be controlled by the magnetic field $h_z$. Figure~\ref{fig4}b shows Fourier spectra $\tilde{m}_{z}(\omega)$ at different $h_z$. The oscillation frequency shifts approximately linearly with $h_z$ from the zero-field value (see the black dashed line at the bottom in Fig.~\ref{fig4}b) with stronger deviations from linearity at  large fields. The amplitude of the oscillations is enhanced (suppressed) when magnetic field is applied towards (away from) the resonance  (cf. Figs.~\ref{fig1} and~\ref{fig4}b).
These magnetic-field dependences are consistent with those found in the conventional central spin problem \cite{BM07}, suggesting the tantalizing possibility to control the electron spin of dense Rydberg gases in an analogous way to solid-state qubits \cite{KA03,*JS03,*CWA04,HR08,WWM10,*TAM12}.  

We emphasize that the defining features of the RCSM 
originate from the unique interplay between the orbital motion and the central spin couplings of environmental atoms. To demonstrate this, in Figs.~\ref{fig2}c,d we plot the results of $A(\omega)$ and $m_z(t)$ in the limiting case of heavy  atoms $m\!\to\!\infty$, where orbital dynamics is completely frozen and the system reduces to the ordinary central spin problem with random couplings. Specifically, for each atomic configuration $\{{\bf r}_1,\ldots,{\bf r}_N\}$, we solve exactly the time evolution of the integrable central spin Hamiltonian $\hat{\bf S}_e\!\cdot\!\sum_{i=1}^N g_i\hat{\bf S}_i$ with the polarized initial condition~\eqref{initial}, obtaining the absorption spectrum via the exact expression $A^{\{{\bf r}_i\}}(\omega)=\sum_{l=1}^{N+1}\delta(\omega-\omega_l)A_l$ with $A_l=1/[1+\sum_i g_i^2/(\omega_l+g_i/2)^2]$. Here, we denote $g_i=g({\bf r}_i)$ and the Bethe roots $\{\omega_l\}$ satisfy 
$\sum_{i=1}^{N}{g_{i}}/(2\omega_l+g_{i})=-1$.
The absorption spectrum is then obtained by taking the average over atomic configurations
\eqn{\label{infmass_abs}
A_{m\to\infty}(\omega)=\sum_{\{{\bf r}_{i}\}}{\rm Prob}[\{{\bf r}_{i}\}]A^{\{{\bf r}_{i}\}}(\omega),
}
where $\rm Prob$ denotes the spatial distribution of environmental atoms determined from the initial wavefunction. 

The results in the infinite-mass limit neither exhibit the characteristic multiple peaks (Fig.~\ref{fig2}c) nor the  oscillations in $m_z(t)$ (Fig.~\ref{fig2}d). The latter is because oscillations cancel out due to the incoherent summation~\eqref{infmass_abs} over the initial distribution while for each realization of atomic positions the central spin still exhibits long-lasting oscillations \cite{BM07}.
These results  demonstrate that the orbital motion of environmental particles, which is absent in the conventional central spin problem, is essential for understanding the  physics of the RCSM.

\begin{figure}[b]
\includegraphics[width=78mm]{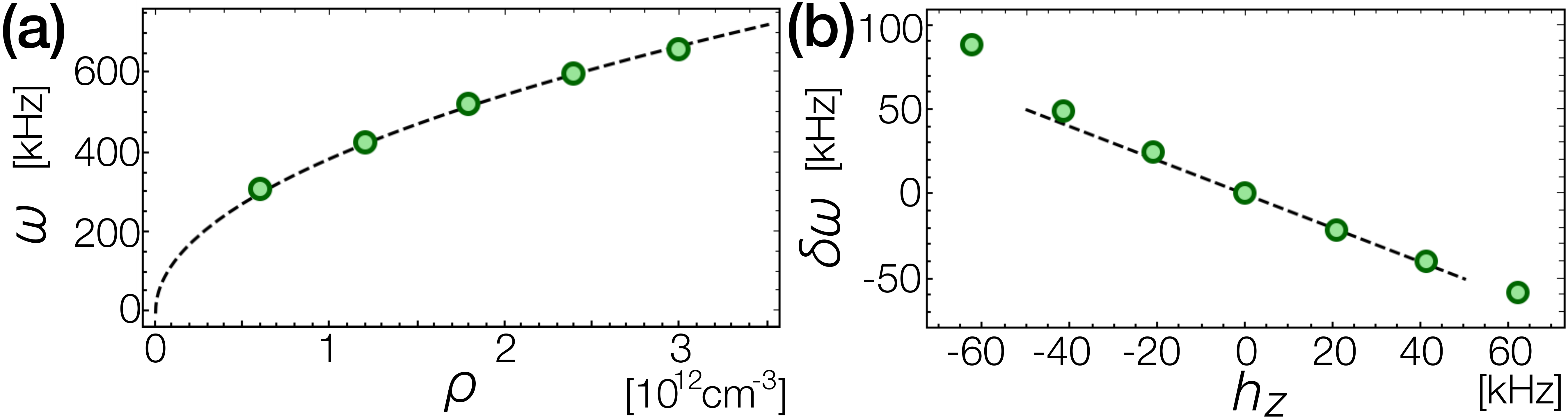} 
\caption{\label{fig5}
The mean frequencies of the central spin dynamics with the anisotropic spin interaction at (a) different densities $\rho$ with zero magnetic field and at (b) different magnetic fields $h_z$ with $\rho=6\times 10^{11}$cm$^{-3}$. The black dashed curve and line indicate the scalings $\omega\propto\sqrt{\rho}$ in (a) and $\delta\omega=-h_z$  in (b), respectively.
}
\end{figure}

{\it Discussions.---}  As a concrete realization, we consider an ensemble of alkaline-earth atoms (e.g., $^{84}$Sr) as a bosonic environment and an alkali atom (e.g., $^{87}$Rb) as a host for the Rydberg excitation. We assume that Sr atoms  have been transferred into the $^{3}P_{1}$ state that is metastable. Effect of the finite lifetime (and thus the nonzero linewidth) can be avoided by  using a Rydberg state with a smaller principal quantum number $n$, leading to a better resolution in spectroscopic measurement. The hyperfine interaction in bath atoms is absent since Sr atoms have no nuclear spins while the Rydberg hyperfine interaction scales as $1/n^3$ and is on the order of $\sim\!\!10{\rm kHz}$. While this energy can be on the scale of molecular binding, it is tiny compared to the spin coupling $g_{\bf r}$. The electron-atom scattering then separates into $J_{\rm tot}\!=\!3/2,1/2$ channels with  corresponding pseudopotentials $V_{3/2,1/2}$. The sign of the scattering length is negative (positive) for the former (latter) channel \cite{Priv}.  
The impurity-boson interaction can be written as $V_{0}+g_{\bf r}\hat{\bf S}_e\cdot\hat{\bf J}_{\bf r}$ with $V_{0}\!=\!(2V_{3/2}\!+\!V_{1/2})/3$ and $g_{\bf r}\!=\!2(V_{3/2}\!-\!V_{1/2})/3$.
 Identifying two internal states of environmental atoms as $|\!\Uparrow\rangle\!=\!|J\!=\!1,J_z\!=\!+1\rangle$ and $|\!\Downarrow\rangle\!=\!|J\!=\!1,J_z\!=\!0\rangle$, we introduce effective spin-1/2 operators by using the correspondence $\hat{J}^{x,y}_{\bf r}\leftrightarrow\sqrt{2}\hat{S}^{x,y}_{\bf r}$ and $\hat{J}^{z}_{\bf r}\leftrightarrow\hat{S}^{z}_{\bf r}+1/2$, leading to the effective Hamiltonian $\hat{H}_{\rm eff}\!=\!\sqrt{2}\hat{H}_{\perp}\!+\!\hat{H}_{\parallel}$. 
 The main difference between this model and the basic RCSM Hamiltonian (\ref{Hamiltonian}) is the anisotropy of the Kondo interaction.  Figure~\ref{fig5} demonstrates that this anisotropy does not  alter our findings qualitatively (see Ref.~\cite{YA19A} for further details).  
 
We note that the present formulation can be applied to analyze a broad class of quantum many-body systems, in which a localized spin is coupled to multiple modes of a many-body environment \cite{YA19A}.  The  spin dynamics found in our analysis suggests an intriguing possibility that  photoexcited electrons can be used to prepare and manipulate mesoscopic spin environments, analogously to what has been demonstrated in solid-state qubits \cite{KA03,*JS03,*CWA04,HR08,WWM10,*TAM12}.
 
\begin{acknowledgments}
We are grateful to Shunsuke Furukawa, Tom Killian, Jesper Levinsen, Meera Parish, Masahito Ueda, and Shuhei Yoshida for fruitful discussions. Y.A. acknowledges support from the Japan Society for the Promotion of Science through Program for Leading Graduate Schools (ALPS) and Grant Nos.~JP16J03613 and JP19K23424, and Harvard University for hospitality. T.S. acknowledges the Thousand-Youth-Talent Program of China. R.S. is supported by the Deutsche Forschungsgemeinschaft (DFG, German Research Foundation) under Germany's Excellence Strategy -- EXC-2111 -- 390814868.
J.I.C. is supported by the ERC QENOCOBA under the EU Horizon 2020 program (grant agreement 742102).
E.D. acknowledges support from Harvard-MIT CUA,  AFOSR Quantum Simulation MURI, 
AFOSR-MURI: Photonic Quantum Matter (award FA95501610323).
\end{acknowledgments}
\bibliography{reference}

\end{document}